\documentclass{IEEEtran}
\usepackage[ddmmyy,nodayofweek]{datetime}
\usepackage{url,todonotes}
\usepackage{times}
\usepackage{graphicx}
\usepackage{rotating}
\usepackage{listings}
\usepackage{listings}
\usepackage[center]{caption}
\usepackage{xspace}

\newcommand{\eg}[0]{\emph{e.g.}\xspace}    % 'e.g. '
\newcommand{\aecute}[0]{{\AE}cute\xspace}
\newcommand\pencil{\textsc{Pencil}\xspace}

\def\OPTL{\textrm{$[$}}
\def\OPTR{\textrm{$]$}}

\definecolor{lightbackground}{rgb}{.98,.98,.97}
\definecolor{darkgray}{rgb}{.3,.3,.3}
\definecolor{darkred}{rgb}{.6,0,0}
\definecolor{darkgreen}{rgb}{0,.6,0}
\definecolor{darkblue}{rgb}{0,0,.6}
\usepackage{listings}
\lstdefinelanguage{pencil}{%
  keywords={[1]for,do,while,if,else,break,continue,return},
  keywords={[2]struct,union,enum,typedef,const,volatile,%
    signed,unsigned,restrict,static,sizeof,typeof,%
    void,char,short,long,int,float,double,boolean,size_t},
  keywords={[3]pragma,pencil,independent,ivdep,reduction,access},
  keywords={[4]inline,noinline,__attribute__,%
    pencil_access,__pencil_kill,KILL,%
    __pencil_maybe,MAYBE,__pencil_use,USE,__pencil_def,DEF,MAY_DEF,%
    PENCIL,__pencil_access,ACCESS,CONST},
  literate={[OPT[}{{\OPTL}}1 {]OPT]}{{\OPTR}}1,
  string=[b]",
  comment=[l]//,
  morecomment=[s]{/*}{*/},
  mathescape=true,
  flexiblecolumns=true,
  tabsize=2,
  captionpos=b,
  frame=single,
  framerule=0pt,
  aboveskip=1pt,
  belowskip=1pt,
  framesep=1pt,
  basicstyle=\ttfamily\footnotesize,
  keywordstyle={[1]\color{darkred}},
  keywordstyle={[2]\color{blue}},
  keywordstyle={[3]\color{darkgreen}\bfseries},
  keywordstyle={[4]\color{darkblue}\bfseries},
  emphstyle=\slshape,
  identifierstyle=\color{black},
  commentstyle=\color{darkgray},
  stringstyle=\color{green}
}

\begin{document}

\title{
\pencil: Towards a Platform-Neutral Compute Intermediate Language for DSLs%
\\ {\small }%\today, \currenttime}%
}

\author{%
\IEEEauthorblockN{Riyadh Baghdadi\IEEEauthorrefmark{2}},
\IEEEauthorblockN{Albert Cohen\IEEEauthorrefmark{2}},
\IEEEauthorblockN{Serge Guelton\IEEEauthorrefmark{2}},
\IEEEauthorblockN{Sven Verdoolaege\IEEEauthorrefmark{2}},
\IEEEauthorblockN{Jun Inoue\IEEEauthorrefmark{2}},
\IEEEauthorblockN{Tobias Grosser\IEEEauthorrefmark{2}},
\newline
\IEEEauthorblockN{Georgia Kouveli\IEEEauthorrefmark{1}},
\IEEEauthorblockN{Alexey Kravets\IEEEauthorrefmark{1}},
\IEEEauthorblockN{Anton Lokhmotov\IEEEauthorrefmark{1}},
\IEEEauthorblockN{Cedric Nugteren\IEEEauthorrefmark{1}},
\IEEEauthorblockN{Fraser Waters\IEEEauthorrefmark{1}\IEEEauthorrefmark{3}},
\newline
\IEEEauthorblockN{Alastair F.~Donaldson\IEEEauthorrefmark{3}}
  \IEEEauthorblockA{\newline\IEEEauthorrefmark{1}
  ARM, Media Processing Division, 110 Fulbourn Road, Cambridge, CB1 9NJ, UK}
  \IEEEauthorblockA{\newline\IEEEauthorrefmark{2}
  INRIA and \'{E}cole Normale Sup\'{e}rieure, D\'{e}partement d'Informatique, 45 Rue d'Ulm, 75005 Paris, France}
  \IEEEauthorblockA{\newline\IEEEauthorrefmark{3}
  Imperial College London, Department of Computing, 180 Queen's Gate, London, SW7 2BZ, UK}

  \thanks{This work was supported by the EU FP7 STEP project CARP (project
  number 287767). The work of T. Grosser was also sponsored in part by a Google Europe Doctoral
  Fellowship in Efficient Computing.}
}

\maketitle

\begin{abstract}

We motivate the design and implementation of a platform-neutral
compute intermediate language (\pencil) for productive and
performance-portable accelerator programming.

\end{abstract}

%-----------------------------------------------------------
\section{Introduction}
\label{sec:intro}

Many systems -- from supercomputer installations to embedded
systems-on-chip -- benefit from using special-purpose
{\em accelerators} which can significantly outperform general-purpose
processors in terms of energy efficiency as well as in terms of
execution speed.

Software for accelerated systems, however, is currently written using
low-level APIs, such as OpenCL and CUDA, which increases the cost of
its development and maintenance.  On the other hand, general-purpose
programming languages like C, C++ and Java do not directly leverage
features of accelerators, such as data-level parallelism, or support
common accelerator programming idioms, such as iteration space tiling.
Furthermore, in many application domains for which accelerators show
promise, such as image processing and computational fluid dynamics, it
is common to program in domain-specific languages (DSLs).

Compiling DSLs directly into OpenCL or CUDA is possible but not
advisable.  For example, to target accelerated platforms effectively
the DSL implementers must develop sophisticated code generation and
optimization techniques.  Given typical budget constraints, they will
likely limit their efforts to a set of techniques useful for a small
number of target platforms (\eg accelerated by NVIDIA GPUs), thus
compromising on performance portability.  Moreover, the implementers
of different DSLs will likely spend their efforts on implementing an
overlapping set of techniques.  Clearly, both teams would benefit if
they could target an efficiently implemented intermediate language.

Beside enhancing productivity, DSLs have the advantage of using high level
constructs that have rich semantics.  These constructs provide a wealth of
information that enable the compiler to optimize and parallelize
the code even for algorithms that are considered to be irregular when
expressed in languages like C.
DSL compilers keep a close control on the generated code, eliminating
many of the problems faced by general-purpose optimizing compilers.

%TODO for IMPACT'13: maybe we should also say that \pencil will help
% easier code verification.  Since the code does not contain pointers.

%TODO: add a complete section about the difference between OpenMP, OpenACC, Fortran and PENCIL.

In this article, we present our work in progress on a platform-neutral
compute intermediate language for DSLs called \pencil.
We give an early overview of \pencil and some of the design guidelines that
will help in its definition.  We show some coding rules, language extensions
and directives that we envisage to include in \pencil along with a preliminary
syntax.  And finally, we present two examples of DSLs and show how they can
be expressed in \pencil.

\section{Overview of \pencil}
\pencil will be a platform-neutral intermediate language for 
multiple high performance DSLs.  An
optimization framework will take care of optimizing and
parallelizing the intermediate language.  In this paper we use the polyhedral
framework~\cite{Fea92} as an example of a
static optimization framework.  The polyhedral framework uses an
algebraic representation and abstraction of programs to reason
about loop transformations, allowing the modeling and application of
complex loop nest transformations addressing most of the parallelism
and locality-enhancing challenges.

The \pencil language is meant to facilitate \emph{automatic} parallelization and
optimization for execution on multi-threaded SIMD hardware; it will thus have
sequential semantics.  The syntax presented in this work is a preliminary
syntax based on C, and benefiting from C99 and the GNU extensions.

\pencil will be suitably high-level to allow straightforward
DSL-to-\pencil compilation, but will provide direct support for common
accelerator features and programming idioms, to allow downstream compilation
into extremely efficient low-level code.
In particular, its features will include extensions and directives (pragmas)
allowing users to supply information about dependences and
memory access patterns that may be difficult or impossible to
analyze automatically, and a low-level API allowing expert programmers to
exert control over performance-related aspects such as scheduling,
vectorization, placement and data layout, when desired.

The information captured by \pencil extensions and directives (pragmas) are similar 
to \aecute metadata~\cite{HowesLDK09}, which have proved successful in
proof-of-concept implementations.  We plan to extend this
initial work in two ways.

First, we will ensure that \pencil can represent both regular and
irregular algorithms suitable for accelerators, by systematically studying
algorithmic `motifs' (originally called `dwarfs') proposed by researchers
from Berkeley~\cite{ViewFromBerkeley}.
\aecute metadata are a close fit for {\em regular} algorithms which typically
have static iteration spaces and memory access patterns, such as dense linear
algebra and stencil computations. 
We have used similar techniques to generate efficient OpenCL code for an
{\em irregular} algorithm -- sparse matrix-vector product -- for several
state-of-the-art sparse matrix formats suited for GPUs~\cite{GreweL11}.

Second, we will investigate the use of directives and extensions in
cross-component optimizations, where dependence information associated with
several computational kernels is collectively exploited to perform
transformations to increase parallelism and locality.
In addition to being useful as a compilation target, \pencil will remain
sufficiently high-level and structured to be used directly as an efficiency
language, particularly for library implementers.
Thus, the cross-component optimizer will be designed to support linking and
transformation of a mixture of \pencil code compiled from DSLs,
hand-written user and library code.

\begin{figure}[h!tb]
 \centering
 \includegraphics[scale=0.50,keepaspectratio=true,bb=50 0 500 822]{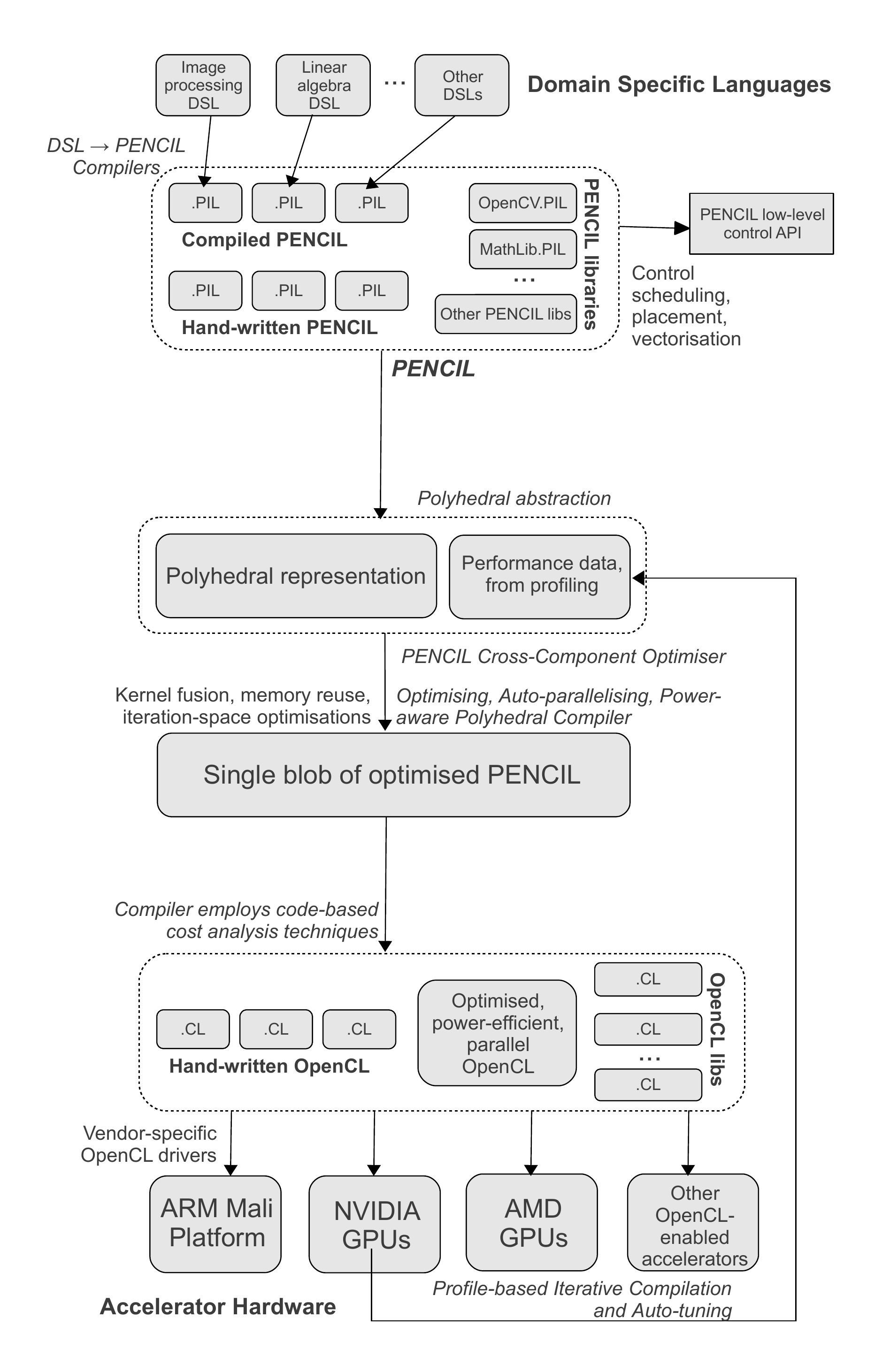}
 \caption{The DSL compilation flow involving \pencil.}
 \label{CARPOverview}
\end{figure}

Figure~\ref{CARPOverview} shows the DSL compilation flow
involving \pencil.
First, a program written in a domain specific
language is translated into \pencil.
The \pencil design aims to make the task of writing a DSL$\rightarrow$\pencil
compiler (the job of the DSL implementer) as straightforward as possible.
Domain specific optimizations are applied during this translation.
Second, the generated \pencil code is combined with hand-written \pencil
codes that implement specific library functions.  This combination of codes is
then optimized and parallelized (using the polyhedral framework
for example).  Finally highly
specialized OpenCL code is generated.  The generated code is tuned
through profile-based iterative compilation and auto-tuning.

%TODO A section to talk about the relation between static
%  optimizations of tasks (intra-task optimization) and dynamic task
%  scheduling (runtime).

%\section{Dynamic task scheduling and static optimizations}
%\pencil should be used to write kernels.  If static scheduling for these
%kernels is not possible due to missing dependence information.

%General execution model: Galois.  The focus of optimizations is intra-task optimizations.
%We do not focus on the runtime system, scheduling of tasks,... We rather focus on intra-task optimizations.

%By looking at Delite/OptiML and the Galois system, it seems that general DSLs require dynamic scheduling of tasks.
%In Galois we have:
%- a task (called operator in Galois).
%- a set of data elements (called active elements in Galois).
%- iterators which apply a task on each one of the data elements:
%    - unordered iterators (for a parallel execution of tasks).
%    - ordered iterators (the order is either specified at the DSL level or
%      at runtime by analyzing dynamic dependences).  In some applications
%      a task working on a node (active element) of the graph may have some
%      dynamic dependences on other nodes.  If two tasks work on the same node,
%      one of them is killed.

%Delite is quite similar.  Dependences for each tasks are first registered.
%The runtime uses these dependences to schedule the tasks.

\subsection{\pencil design}
In order to guarantee the correctness of optimizations,
compilers usually take conservative assumptions.  These
conservative assumptions reduce the ability of the compiler
to find optimizations.  The compiler may assume,
for example, that two
pointers may alias, whereas the pointers do not actually alias.
The fact that the two pointers do
not alias is in general well known to the programmer and to the
DSL compiler, but this information is not transmitted, in general,
to the compiler.

To address this problem, \pencil sets coding rules that
may be used by DSL compilers and by expert \pencil programmers
in order to enhance the ability of the compiler to perform
static code analysis.  Some of these rules will be checked and
enforced by the \pencil compiler, while some others are left up to
the programmer or DSL compiler.

The current syntax of \pencil uses C annotations and extensions
where possible.  As such, a \pencil program, in the current state,
retains the standard syntax and semantics of a C program and can be processed
by an ordinary C compiler.  Semantical additions to C make use of custom
GNU extensions and directives.

While designing \pencil, we are putting a strong emphasis
on the definition of annotation syntaxes and coding rules that may be easily
lowered to compiler intermediate representations using attributes and
built-in functions, mainly because we are considering an equivalent
LLVM IR syntax for \pencil.

\subsection{Examples of \pencil coding rules, extensions and directives}

\subsubsection{Coding rules}
One of the main characteristics of \pencil{} is its
restriction on pointer usage in order to eliminate aliasing.
\pencil will accept only non-array variables to be passed by pointer.
Array parameters must be passed using the C99 VLA syntax and
must be qualified \lstinline!restrict!, \lstinline!const!, and
\lstinline!static! with the same syntax and semantics as in C99.  For example:

\begin{lstlisting}
/* The following function is PENCIL-compliant.  */
void foo (int a[restrict const static 5]) {
  /* `a[]' is const but its elements are not.  */
  a[0] = 1;
  
  /* Here const is not required.  Local variables are
     not coerced to pointers.  */
  int c[2];
}

/* Example of non PENCIL-compliant declarations.  */
void bar (int *(d[4])) {
  int *e;
}
\end{lstlisting}

%TODO: Jun rephrase pointer manipulation.  see the comment of Ally.

Readers may recall that C99 coerces the type of \lstinline!a! in function
\lstinline!foo! to \lstinline!int*!, but we require the explicit array
declaration syntax to reinforce that \pencil{} disallows pointer arithmetic.
We may also find ways to leverage the declared array size information 
in \pencil{} compilers in the future.

A pass-by-pointer parameter should be declared in the receiving function’s
prototype as a \lstinline!const restrict! pointer.  %  In the body of the receiving 
% function such a parameter can only be dereferenced using the 
% \lstinline!*! operator.  The \lstinline![]! notation will be reserved
% exclusively for parameters, variables, and struct members declared as arrays.
These restrictions guarantee that a pointer can only point to a fixed memory
region throughout its lifetime and that different pointers never point to the
same memory region.

Other coding rules that we envisage to enforce in \pencil programs include
the constraint
that recursion (whether direct or indirect) and unstructured control flow 
(via \lstinline!goto!s) are not allowed.

\subsubsection{Extensions}

\pencil provides \emph{access summary functions} for describing the
data access patterns of a function.  This mechanism may be applied to any
function, including those whose behaviors are too complex for the compiler to
infer accurately, as well as library functions whose source code is not
available to the \pencil compiler and/or which internally uses features of
C that are banned in \pencil.  In the following example, \lstinline!ACCESS!
declares that \lstinline!foo!  performs the same data access as
\lstinline!foo_summary!  (array qualifiers are omitted for brevity):

%TODO: clarify, the for loop in the summary is not the for loop in the original code,
% it's just a way to specify the iteration domain.
% Jeron said: I thought only the computation is abstracted. It would be good to state that both computation and
%control flow are abstracted from in the summary and that the for loop
%in the example is simply notation for quantification over all n (a specification for the iteration domain), not
%related to the loop in the foo-function.

\begin{lstlisting}
void foo_summary(int n, int A[n], int B[n],
                       int C[n])
{
  for (int i=0; i<n; i++) {
    DEF(A[i]); USE(B[i]); MAY_DEF(B[i]);
  } 
  if (n < 4) DEF(C[0]); // one-element def
  USE(A[n-1]);
}

void foo(int n, int A[n], int B[n], int C[n])
              ACCESS(foo_summary(n, A, B, C))
{

  int i;
  for (i=0; i<n; i++) {
    A[i] = B[i];
    B[rand() % n] = 42;
  }
  if (n < 4) C[0] = A[n-1];
}
\end{lstlisting}

The macros \lstinline!DEF!, \lstinline!USE!, and \lstinline!MAY_DEF!  expand to
built-in functions that modify, use, or may modify their argument, respectively,
but which are guaranteed not to be accidentally optimized out in upstream
compiler passes.
%A summary function can contain any \pencil code, and
%everything except the control flow and calls to macros such as \lstinline!DEF!
%and \lstinline!USE! is ignored.
The actual accesses summarized by the function are defined by the
array elements traversed along the execution of the summary
function. Control flow and C instructions are only meant to drive the
enumeration of these accesses. Since these summaries are meant to be
processed by a static analyzer, non-affine control flow may lead to
further discrepancies between may-write and must-write access sets.
For example, the result of such a static analysis could take the form
of three distinct access relations, mapping each iteration of the
summarized function call and/or its parameters to a set of may-write,
must-write, and read accesses respectively.

\subsubsection{Directives}

\pencil uses directives inspired by OpenMP, OpenACC and advanced
vectorizing compilers.

%A \emph{\pencil region} denotes a C function or single-entry single-exit
%control-flow region obeying the \pencil coding rules.\footnote{It is
%possible to identify such \pencil regions using a specific syntax.
%A compiler may also extract \pencil regions automatically.}
%\pencil regions are marked using the \lstinline!pencil! directive:

%\begin{lstlisting}
%#pragma pencil
%{
  %\ldots$ /* Hot piece of code.  */
%}
%\end{lstlisting}

The restrictions presented in the previous section simplify data and
control dependence analysis, which gives \pencil compilers a boost
in loop optimizations.  When this falls short
of providing the compiler with necessary static information, however,
dependence information can be explicitly supplied as directives.  One
such directive is
\begin{lstlisting}
#pragma pencil independent [OPT[($l_1, \ldots, l_n$)]OPT]
\end{lstlisting}
The list $l_1, \ldots, l_n$
indicates the labeled statements on which the loop independence is
guaranteed.  A
statement that appears in an \lstinline!independent! clause is assumed not to
have any loop-carried dependence with any other statement in the loop.  If this
list is omitted then all statements in the loop body are free of dependences
carried by the annotated loop.  In the following example:
\begin{lstlisting}
#pragma pencil independent
for (int i=0; i<N; i++)
  A[t[i]]++;
\end{lstlisting}
different iterations of the loop may write to the
same array location. The write location depends on the value of
\lstinline!t[i]!. In order to parallelize the loop, the compiler
needs to make sure that there is no loop-carried dependence, but
proving this property is not possible at compile time.  Thus, the
compiler considers conservatively that there may be a dependence
between the different iterations and the loop is not
parallelized. If the DSL compiler or the expert \pencil programmer
know that all values of \lstinline!t[i]! are different then she
should insert an \lstinline!independent! pragma to indicate that different
iterations of the loop are independent. This will not only enable the
parallelization of the loop, but also provide valuable static
information to other loop transformations and optimizations.

% In the following example:
% \begin{lstlisting}
% #pragma pencil independent(s1)
% for (int i=0; i<N; i++) {
%   $s_1$: A[t[i]]++;
%   $s_2$: B[i]=A[x[i]];
%   $s_3$: B[y[i]]++;
% }
% \end{lstlisting}
% there is no loop-carried dependence between the
% different instances of $s_1$ ($s_1$ in different iterations), there is
% no loop-carried dependence between $s_1$ and $s_2$ and also no
% loop-carried dependence between $s_1$ and $s_3$. But, the
% independent pragma doesn't guarantee that there is no loop-carried
% dependences between $s_2$ and $s_3$.

% indicates that a reduction operation is performed on the
% scalars that appear in its list. The polyhedral framework will
% use this information and parallelize the loop accordingly. The
% syntax and the semantics of this pragma are equivalent to
% the syntax and semantics of the reduction clause defined in
% OpenMP and OpenACC.

Unlike the OpenMP \lstinline!parallel! pragma, it is possible to use
the \lstinline!independent! pragma on while loops to indicate
that there is no dependence between the different iterations
of the while loop.  It is up to the compiler to use this information
to optimize the code.  Moreover, the \lstinline!independent! pragma allows fine grain
code description as its scope may be limited to only one statement in the
loop body.

\pencil also defines a reduction directive equivalent to the
reduction directive defined in OpenMP and OpenACC.  It has the
following syntax:
\begin{lstlisting}
#pragma pencil reduction (operator : scalars)
\end{lstlisting}

Note that \pencil does not compete with OpenMP, actually \pencil
complements OpenMP and, in general, the coding rules defined by \pencil
are useful for compiler optimizations even if they are
used outside \pencil.

% Thanks to the restrictions on pointer usage, it is possible to identify
% automatically in the optimization framework any reduction operation on
% scalars and thus there is no need to use a reduction pragma as happens in
% OpenMP.  The general pattern of reductions that can be identified
% automatically is

% \begin{lstlisting}
%   scalar = scalar op expression
% \end{lstlisting}
% where $op$ is an associative and commutative operator.\footnote{We are
% considering to generalise $op$ to include any commutative and associative
% function marked as such by the user through function attributes that we
% may introduce.  This is similar to the extension suggested
% by~\cite{duran_proposal_2010}.}  The polyhedral framework will consider
% that the loop nest contains a reduction and will optimize and parallelize
% the loop accordingly.

All in all, this feature set provides a language whose overall semantics is
sequential but which places conventions and restrictions that increase the
static information available to the compiler, thus enabling the compiler to do
more aggressive loop nest optimizations and parallelization.  Although the
language is sequential, the information about parallelism available at the
DSL level is not lost, because this information is expressed in \pencil
through directives like \lstinline!independent! which indicates the absence of
dependences in a given loop.  Any lower level compiler can use this information
not only to parallelize the loop but to apply other optimizations as well.
Expressing the absence of dependences is more powerful than expressing
only parallelism.

\section{Examples of DSL translation into \pencil}

This section provides examples of DSLs that can be mapped into
\pencil, and benefit from the optimizations provided by \pencil
compilers, including polyhedral compilation methods. Some DSLs are
mostly designed for programmer productivity, and their compilation
flow typically combines specific passes for abstraction penalty
removal with more generic optimization passes. Such DSLs should
immediately benefit from \pencil with minor modifications to their
compilation flow. Other DSLs involve a lot of domain-specific
information available for compile-time optimizations. Since a large
number of these optimizations are actually generic ones, expressible
as loop transformations and storage mapping choices, \pencil will also
contribute to the simplification of their tool flow.

In all of the following examples, memory access
information should be used to annotate functions called from within
the kernels.  Moreover, the coding rules are mandatory
to enable a precise dependence analysis.

\subsection{OP2 library}

OP2 \cite{OP2} is a state-of-the-art library for parallelizing unstructured
mesh computations.  It restricts the computational kernel's data-access
pattern, simplifying dependence analyses and facilitating task decomposition,
scheduling, and data layout.  While a great deal of OP2's innovations lies in
its efficient backend implementations, it is noteworthy how \pencil
captures OP2's most important restrictions.  Let us illustrate
this with the following program using OP2's C++ binding, adapted from
\cite{OP2}.  Functions named with the \verb|op_| prefix constitute OP2's API.
For the sake of conciseness we have omitted string parameters that are
used for dynamic type checking and diagnostics.

\begin{lstlisting}
void kernel (double *edge,
                 double *cell0, double *cell1)
{
  *cell1 += *edge; *cell0 += *edge;
}

void main_loop (int ncells, int nedges,
                     int *edge_to_cells,
                     double *edge_data,
                     double *cell_data)
{
  op_set cells = op_decl_set (ncells);
  op_set edges = op_decl_set (nedges);
  op_map pecell = op_decl_map (edges, cells, 2,
                                        edge_to_cells);
  op_dat dcells = op_decl_dat (cells, 1, cell_data);
  op_dat dedges = op_decl_dat (edges, 1, edge_data);

  op_par_loop (kernel, edges,
    op_arg(dedges, -1, OP_ID, 1, OP_READ),
    op_arg(dcells, 0, pecell, 1, OP_INC),
    op_arg(dcells, 1, pecell, 1, OP_INC));
}
\end{lstlisting}

In this example, we assume a 2D mesh with \verb|ncells| cells, numbered (or
indexed) from \verb|0| through \verb|ncells-1|, and a total of \verb|nedges|
edges also numbered from \verb|0|.  We ignore boundary edges for simplicity and
assume that each edge falls between exactly two cells.  The input
\verb|edge_to_cells| is a 1-to-2 mapping that indicates which edge touches which cells --
the edge with index \verb|i| touches the cells with indices
\verb|edge_to_cells[2*i]| and \verb|edge_to_cells[2*i+1]|.  Every edge or cell
carries one double-precision floating point data, specified by \verb|edge_data|
or \verb|cell_data|, respectively.  The main computational kernel adds to each
cell all data coming in from its edges; we wish to do this for all cells.

The first six lines of \verb|main_loop()| just communicate this setup to OP2.
\verb|op_decl_set()| is used to declare the set of cells and the set of edges,
while \verb|op_decl_map()| defines the relationship between them using
\verb|edge_to_cells|.  The argument \verb|2| indicates to OP2 that this is a 1-to-2
mapping.  Conceptually \verb|pecell| is just a copy of \verb|edge_to_cells|,
made opaque so that OP2 is not constrained by the layout or location of
\verb|edge_to_cells|.  Finally \verb|op_dec_dat()| attaches data to the cells
and edges.

The most interesting part is \verb|op_par_loop()|, which is conceptually
equivalent to the following plain C loop:
\begin{lstlisting}
  for (int i = 0; i < nedges; ++i)
    kernel (&dedges[i],
              &dcells[pecell[2*i]],
              &dcells[pecell[2*i+1]]);
\end{lstlisting}
In words, \verb|op_par_loop()| iterates over the indices of \verb|edges|,
calling a kernel on the data associated with each index.  The three calls to
\verb|op_arg()| are used to indicate the arguments of \verb|kernel()|, and to
describe how each argument is being accessed.  

For example
\begin{lstlisting}
  op_arg(dcells, 0, pecell, 1, OP_INC)
\end{lstlisting}
tells \verb|op_par_loop()| to that the first argument of \verb|kernel| is
\verb|dcells|; and that the index used to access \verb|dcells| is calculated
by looking up \verb|pecell| at the loop
index \verb|i| and adding the offset \verb|0|; the number of data elements
passed to the kernel is $1$ starting at the translated index.  \verb|OP_ID|
is used to indicate that the loop index should be used directly to address
the data.  The last argument \verb|OP_INC| is a hint on how the kernel function
accesses this data; it means the data is the subject of a global-reduction sum,
as seen for \verb|cell0| and \verb|cell1| in the above example.  The other
possible hints are \verb|OP_READ|, \verb|OP_WRITE|, and \verb|OP_RW|.  For the
last two, the kernel code must ensure that no data conflict is possible between
different iterations.

It should be clear that OP2's semantics is correctly captured in \pencil by
translation to a \verb|for| loop like the one above, with the caveat that
\verb|kernel| must be either inlined or modified to
\begin{lstlisting}
void kernel (double edge[], int ie,
                 double cell0[], int i0,
                 double cell1[], int i1)
{ cell1[i1] += edge[ie];
  cell0[i0] += edge[ie]; }
\end{lstlisting}
(because \pencil does not allow pointers).  The translated \verb|for| loop is
legal \pencil.
%, though perhaps the \verb|kernel| must be treated as an external
%non-\pencil function.
The other parts -- the first six lines of
\verb|main_loop()| -- simply reify and constrain the input data, which is
unnecessary in \pencil.

This is not surprising, as OP2 is a more aggressively restricted DSL than
\pencil.  The more interesting fact is how much of OP2's static information can
be captured in \pencil.  The single greatest benefit from OP2's programming model
is probably elimination of pointer analysis.  This is built into \pencil.  Of
OP2's access hints, \verb|OP_INC| can be expressed with a \verb|reduction|
pragma, the conflict-freedom requirement of \verb|OP_WRITE|/\verb|OP_RW| can be
expressed with \verb|#pragma independent|, and \verb|OP_READ| should be
inferable from the source code.

One aspect of OP2 that is not currently captured explicitly by \pencil is
allowing the un-associativity of floating point arithmetic to compromise
bit-wise reproducibility of results.  The example program above suffers from
the problem that parallelizing the loop in any way compromises numerical
precision to some extent.  This is a long-standing and well-known issue, often
handled by providing a switch or pragma to allow trading precision for
efficiency.  We plan to follow this well-accepted practice.

\subsection{Delite/OptiML}

OptiML~\cite{sujeeth_optiml_2011} is a DSL for
machine learning built on top of Delite~\cite{chafi_domain-specific_2011},
a framework for creating implicitly parallel DSLs.

An OptiML program is actually a program generator embedded in Scala.
It uses meta-programming to construct a symbolic representation of the
DSL program as it is executed.  Each program expression, such as
\lstinline!if(c) a else b!, constructs an IR node when the program is
run.  Instead of using a control flow graph (CFG) for the different
statements with fixed basic blocks, Delite uses
a "sea of nodes" \cite{brown_heterogeneous_2011} as an IR representation.
Nodes are connected with respect to their (input and control) dependences but
are allowed to float freely otherwise.

The Delite IR provides several operators.  A given DSL may use a subset
of these operators and may also extend existing operators to create new
ones.

OptiML programs operate on the high-level mutable types Vector[T] and
Matrix[T] and provides $4$ main control 
structures: \lstinline!sum!, vector construction,
\lstinline!untilconverged! and \lstinline!gradient!.
We enumerate these structures and show how they can be mapped to \pencil.
 
 \begin{itemize}
  \item \lstinline!sum!: expresses generic summations over an indexed range.
  It calculates $\sum f(i)$ where $f(i)$ is a user-defined function.
  For example
  \begin{lstlisting}
    val x = sum(0,100) { i => exp(i) }
  \end{lstlisting}
  calculates
  \begin{lstlisting}
    x = exp(0) + exp(1) + exp(2) + $\dots$   
  \end{lstlisting}
  \lstinline!sum! is implemented as a parallel tree-reduce and can be translated
  into \pencil using a \lstinline!for! loop and a \lstinline!reduction! directive.
% the reduction should
% be identified automatically by the underlying optimization framework:
  \begin{lstlisting}
  x = exp(0)
  #pragma pencil reduction (+:x)
  for (i=1; i<=100; i++)
    x += exp(i);
  \end{lstlisting}
  
  \item vector construction: implemented as a parallel map in Delite.
  It has the following form 
  \begin{lstlisting}
    val my_vector = (0::end) { i => 0 }
  \end{lstlisting}
  and can be translated into \pencil using a simple \lstinline!for! loop.  There
  is no need in this case to use the \lstinline!independent! pragma, as the loop
  nest is always affine and the underlying optimization tools that operate on
  \pencil will be able to recover the parallelism and generate a parallel code.
  \begin{lstlisting}
  for (i=0; i<=end; i++)
    my_vector[i] = 0;
  \end{lstlisting}
  
 \item \lstinline!untilconverged!: an iterative control structure that iterates
 until reaching a convergence criterion. Each iteration produces
 a value, and the loop converges when the difference between
 values in consecutive iterations falls below a supplied threshold.
 This control structure is implemented in \pencil as a sequential
 loop.
 
%   In some iterative applications, there is a limited data parallelism.
%   However, when the algorithm is robust to minor perturbations,
%   OptiML designers suggest that dependences can be relaxed to provide
%   more task parallelism at the cost of a marginal loss in accuracy.
%   If \emph{untilconverged} is relaxed, a given number of iterations (indicated
%   at the level of OptiML as a percentage of iterations) are allowed to run in
%   parallel and race. \pencil allows the possibility of a parallel execution
%   of those iterations identified by OptiML by ignoring the
%   dependences between those iterations through the \lstinline!independent! pragma.
%   The loop generated by the DSL compiler in this case needs to be strip-mined in
%   order to allow for the parallelization.
 
 \item gradient descent: is a specialized version of \emph{untilconverged}
 that implements the \emph{gradient} descent algorithm for exponential family
 models. It provides batch and stochastic variants.  The batch variant
 uses a parallel algorithm and thus it can be mapped, in \pencil,
 into a \lstinline!for! loop annotated with the \lstinline!independent!
 pragma to indicate that there is no loop carried dependence.
 The stochastic variant is mapped into a sequential \lstinline!for! loop
 as the algorithm is not parallel.
 \end{itemize}

%----------------------------------------------------------
\section{Conclusion}
\label{sec:conclusion}

% We are considering an equivalent LLVM IR syntax for \pencil, whose
% specification will naturally derive from the source-level one.

We proposed \pencil{}, a platform-neutral compute intermediate
language for productive and performance-portable accelerator
programming. This intermediate language facilitates the design and
implementation of high-level programming environments for parallel
architectures. In particular, we believe \pencil{} reduces the
complexity and costs of exploiting heterogeneous systems.

\paragraph*{Acknowledgments}

This work was partly supported by the European FP7 project CARP
id.\ 287767.

%==========================================================
%  Bibliography
%==========================================================
\bibliographystyle{plain}
\bibliography{wolfhpc}

\end{document}